# Materials Informatics for Heat Transfer: Recent Progresses and Perspectives

Shenghong Ju[1,2] and Junichiro Shiomi[1,2,3]*

[1]Department of Mechanical Engineering, The University of Tokyo, 7-3-1 Hongo, Bunkyo, Tokyo 113-8656, Japan

[2]Center for Materials research by Information Integration (CMI2), Research and Services Division of Materials Data and Integrated System (MaDIS), National Institute for Materials Science (NIMS), 1-2-1 Sengen, Tsukuba, Ibaraki 305-0047, Japan

[3]Core Research for Evolutional Science and Technology (CREST), Japan Science and Technology Agency (JST), 4-1-8, Kawaguchi, Saitama 332-0012, Japan

*Corresponding email: shiomi@photon.t.u-tokyo.ac.jp

**ABSTRACT**

With the advances in materials and integration of electronics and thermoelectrics, the demand for novel crystalline materials with ultimate high/low thermal conductivity is increasing. However, search for optimal thermal materials is challenge due to the tremendous degrees of freedom in the composition and structure of crystal compounds and nanostructures, and thus empirical search would be exhausting. Materials informatics, which combines the simulation/experiment with machine learning, is now gaining great attention as a tool to accelerate the search of novel thermal materials. In this review, we discuss recent progress in developing materials informatics for heat transport: the exploration of crystals with high/low thermal conductivity via high-throughput screening, and nanostructure design for high/low thermal conductance using the Bayesian optimization and Monte Carlo tree search. The progresses show that the materials informatics method are useful for designing thermal functional materials. We end by addressing the remaining issues and challenges for further development.

## 1. Introduction

Heat transfer plays an important role in thermal management applications such as heat exchangers, thermal interface materials, heat pipes, heat radiators, thermoelectrics, thermal barrier coating, and thermal insulators [1-4]. Exploration and designing of materials and structures with desired thermal transport properties have large potential for application. However, two bottlenecks limit the designing efficiency: materials selection and structure designing. Selecting the most appropriate material from a large number of candidates is the first key question to face during the designing of thermal devices. Currently, databases including tens of thousands of crystal compounds have been constructed, including Materials Project [5], AFLOW [6], ICSD [7], OQMD [8,9], and AtomWork [10], as shown in Fig. 1 (a). The thermal property of materials varies in a wide range. Taking thermal conductivity as an example, the order ranges from hundredths of $Wm^{-1}K^{-1}$ to thousands of $Wm^{-1}K^{-1}$. Discovery of materials with very low or high thermal conductivity remains an experimental challenge due to high cost and time-consuming synthesis procedures. The other bottleneck is perhaps more challenging. As the length scale of materials decreases to nanoscale, heat conduction becomes more controllable through manipulating the nanostructures, as shown in Fig. 1 (b). Due to the various choices of structure parameters and coupled effects, it is difficult to quickly obtain the optimal nanostructure with desired thermal property from tremendous number of candidates. If exploring the materials or structures one by one using traditional heat transfer analysis method, it will become time-consuming and low-efficiency.

The key next-generation technology to solve the above bottlenecks is materials informatics (MI): integration of material property calculations or measurements with informatics method to accelerate the material discovery and design [11-13]. During the past decade, MI has been successfully applied to design cathode materials of the lithium-ion battery, drugs, polymers, catalysis [14-18], and many others. The application of MI on thermal transport has also been gradually developed. In this review, we summarize the most recent progress of the MI application in heat transfer field. The review is organized as follows. In the first part, we summarize the recent progress of high-throughput screening for ultimate high/low lattice thermal conductivity materials. In the second part, we introduce the nanostructure designing/optimization with maximum/minimum thermal conductance using Bayesian optimization and Monte Carlo Tree search. We hope this review will provide useful guidance for extending the application of MI in the heat transfer field.

## 2. High-throughput screening

High-throughput screening (HTS) is a combination of machine-learning algorithms, physical insights, and automatic ab-initio calculations, which can considerably speed up the selection of best materials from databases for a given objective and has been successfully applied in many fields including catalysis [17,18], battery technologies [16], thermoelectric materials [19,20], chemical probes [21], polymers [22,23] and magnetic materials [24]. In this section, we will introduce recent progress of HTS in the field of heat transfer aiming for high performance thermoelectric materials with low thermal conductivity [19,20,25,26].

Application of machine learning requires data, descriptors, and machine-learning models. The data for heat transfer can come from either or both calculations and experiments. Although there are material-property databases such as Materials Project [5], AFLOW [6], ICSD [7], OQMD [8,9], etc., with thousands of entries for formation enthalpy, bandgaps, modulus, the ones of thermal properties have been limited due to relative difficulty and complexity of calculating or measuring thermal properties particularly thermal transport properties. However, over the past decades, the anharmonic lattice dynamics (ALD) method using interatomic force constants (IFCs) obtained by first-principles has been developed to accurately calculate thermal conductivity [27-29], and the thermal-property data are becoming more accessible. The data involved in the ALD calculation can be divided into 3 types as shown in Fig. 2: (I) general properties associated with the crystals, (II) harmonic properties associated with harmonic IFCs, and (III) anharmonic properties associated with anharmonic IFCs. The type-I data include lattice parameters, atom coordinates, number of elements, number of sites, density, volume, volume/atom, spacegroup number (symmetry), atom types, atom numbers, etc., and all of the information can be obtained directly from crystal databases. The type-II data include phonon dispersion relations, phonon density of states, and group velocity. The type-III data include phonon relaxation time, mean free path, and Grüneisen parameter. For instance, a harmonic phonon database that covers about 10000 crystals has been built up by Togo et al [30], and we expect more to come in near future. With type-II and type-III data, the value of LTC can be calculated by solving the steady-state Boltzmann transport equation (BTE),

$$-\boldsymbol{v}_{qj} \cdot \nabla T \left( \frac{\partial n_{qj}}{\partial T} \right) + \left. \frac{\partial n_{qj}}{\partial t} \right|_{scattering} = 0, \tag{1}$$

where $n$ is the phonon distribution function, $\boldsymbol{qj}$ is the phonon mode, and $\boldsymbol{v}$ is the group velocity. Under relaxation time approximation (RTA) [27,31,32], the LTC can be calculated by,

$$\kappa_{\alpha\beta} = \frac{1}{\Omega N_q} \sum_{q,j} c_{qj} v_{qj}^{\alpha} v_{qj}^{\beta} \tau_{qj}, \tag{2}$$

where $\Omega$ is the volume of the primitive unit cell, $N_q$ is the number of $\boldsymbol{q}$ points, $\alpha$ and $\beta$ indicate the velocity components, $c_{qj}$, $v_{qj}$ and $\tau_{qj}$ are heat capacity, group velocity and relaxation time.

The general strategy in HTS is to incorporate every information of the crystal that may have correlation with the objective thermal properties, and the actual implementation depends on the availability of the data. Carrete et al. [25] screened 79,057 half-Heusler compounds to find mechanically stable semiconductors with low LTC. After removing compounds with positive formation enthalpies and zero band gap, harmonic calculation was carried out for the remaining 995 compounds with lowest-enthalpy configurations, and 450 mechanically stable semiconductors were further selected. Note that the local-density approximations (LDA) and generalized gradient approximation (GGA) theory used in the work typically underestimate the bandgap. The AFLOWLIB database [6] was then used to test the stability based on the convex hull of the ternary phase diagrams, and this finally gave 75 thermodynamically stable compounds. The found lowest thermal conductivity compounds are PtLaSb, RhLaTe, and SbNaSr, with thermal conductivity of 1.72 $Wm^{-1}K^{-1}$, 2.84 $Wm^{-1}K^{-1}$, and 3.49 $Wm^{-1}K^{-1}$, respectively. Roekeghem et al. [26] performed the screening of mechanical stable compounds of oxide and fluoride perovskites at high temperatures using finite-temperature phonon calculations. They found that the thermal conductivity of fluorides are generally lower than oxides largely due to a lower ionic charge. Wang et al. [19] screened several thousand compounds from the ICSD database [7] and gave guiding rules for searching for better thermoelectric materials according to the correlations between the power factor and different physical properties, for example, sintered thermoelectric compounds with large band gaps, heavy carrier effective masses, and more atoms per primitive cell are expected to have large power factors.

When screening materials using machine learning, the selection of descriptors is important and challenging. While, in theory, all the three types of data involved in ALD calculation can be used as descriptors the structure descriptors or/and elemental descriptors are often adopted as they are largely accessible. The structure descriptors of crystals include mainly the lattice parameters, number of elements, number of phonon branches, number of sites, density, volume, volume/atom, energy, energy/atom, and spacegroup number. Besides the traditional structure descriptors, the angular distribution function, radial distribution function [33] and X-ray Diffraction data (XRD) [34] are also potential descriptors. The elemental descriptors are the atomic number, atomic weight, empirical covalent atomic radius, position in periodic table, Pettifor chemical scale, Pauling electronegativity. It was suggested that combinations of different types of descriptors can be useful for machine learning [33]. G. Slack [35] proposed four important descriptors for finding crystals with high thermal conductivity: low atomic mass, strong bonding, simple crystal structure, and low anharmonicity. Carrete et al. [25] studied three types of descriptors correlated with thermal conductivity extensively: chemical information, general crystal information and specific thermal conductivity information. Their result indicates that using only chemical information descriptor can achieve a sufficient prediction. By analyzing the importance of variables in the classification, it was identified that a low Pettifor scale and a large average Pauling electronegativity are the most important descriptors for obtaining low LTC. It was also found the lattice constant is correlated with the sum of atomic radii in the compounds and the materials whose elements with large atomic radii have high possibility of lower LTC. It is easy to understand since large atomic radii generally means heavy atoms, leading to small group velocity, larger lattice constant, and small heat capacity. In the work by Seko et al. [20], even though the rocksalt PbSe was found efficiently with only two descriptors (volume and density), the searching of the third-lowest rocksalt LiI took 65 observations, which is worse than the random search and indicates that using only two descriptors is not sufficient and robust for the screening. To overcome this problem, the descriptors of 34 elements involved in the 101 compounds were added using a set of binary digits representing the presence of chemical elements. The authors stated that the use of the elemental descriptors was found to improve the robustness of the efficient search.

Various models including elastic net [36], support vector regression [37], bagging (bootstrap aggregating) [38], random forest [39], gradient boosting for regression [40], artificial neural network [41], Gaussian process regression [42], clustering algorithms [43,44], transfer learning et al., are available to build up the prediction models based on collected data. Carrete et al. [25] performed the transfer prediction and the random forest regression based on 32 fully calculated ALD cases. According to empirical observation, the IFCs show a high degree of transferability among compounds with the same crystal structure [29]. This means a single set of anharmonic IFCs could be used to estimate LTC of series of materials. The comparison between LTC from accurate ALD calculation and transfer prediction shows a Spearman rank correlation coefficient (which measures the strength of association between two variables) of 0.93, and this indicates the effectiveness of transfer prediction. The random forest model avoids the extreme predictions with nonphysical magnitudes, resulting in a narrow distribution than that of transfer prediction. The bimodal-shape distribution of random forest also suggests that two groups of half-Heuslers with different LTC can be classified. Seko et al. [20] calculated the LTC of 101 compounds with rocksalt, zincblende and wurtzite types of structures using the ALD calculation. They adopted the kriging method based on the Gaussian process regression to build up a prediction model using the volume and density information. The kriging search for the lowest LTC compounds among calculated 101 compounds is faster (required calculation of less number of compounds) than the random search. Based on the prediction model built by the volume, density and 34 elemental descriptors for the 101 LTC data, a ranking for low-LTC compounds is evaluated among 54,779 compounds. 221 compounds are expected to show lower LTC than that of rocksalt PbSe at room temperature. Their distribution in volume-density space is in a wide range which indicates it is difficult to pick them up without performing the kriging search. The top eight compounds were calculated with accurate ALD calculation and five of them (RbPbI$_3$, PbIBr, PbRb$_4$Br$_6$, PbICl, PbClBr) give LTC lower than 0.2 Wm$^{-1}$K$^{-1}$ at room temperature, which confirms the powerfulness of the prediction model based on Gaussian process regression for high-efficient discovering of low LTC compounds. While the database-screening work related with the thermal conductivity reported so far have focused mainly on the low lattice thermal conductivity crystals, we have recently performed the hierarchical screening and transfer learning screening of the high lattice thermal conductivity materials [45], which will be published soon.

## 3. Structure designing/optimization

As the length scale of semiconductor materials reaches nanoscale, the transport of main heat carriers (phonons) becomes more ballistic, which makes it possible to tune the phonon transport by manipulating the nanostructures. However, it is rather difficult to identify the detail optimal structure for phonon transport due to the various and coupled parameters including roughness [46,47], vacancy defects [48], lattice orientation [49,50], nanoinclusions [51], and interfacial adhesion or bonding [52,53]. Besides, the coupling of constructive/deconstructive phonon interference and resonance effects in superlattices [54-57], nanocrystals [58], nanocomposites [59] makes the structure designing and optimization more complicated. Developing an effective optimization method for designing nanostructures with desired thermal property is necessary and has great potential for application. In this part, we summarize recent work on building connections and feedback between the traditional atomistic Green's function (AGF) method [46,60] and the informatics methods: Bayesian optimization [61] and Monte Carlo tree search [62].

### 3.1 Bayesian optimization

Bayesian optimization (BO) is an experimental design algorithm based on machine learning. The main processes of using BO is shown in Fig. 3. Suppose that the transport property of *n* candidate structures are initially calculated, and we are to choose the next one to calculate. The Bayesian linear regression model with random feature map is used to build the prediction model based on the *n* pairs of structures and calculated transport properties,

$$y = \mathbf{w}^T \varphi(\mathbf{x}) + \varepsilon, \tag{3}$$

where $\mathbf{x}$ is a *d*-dimensional descriptor vector corresponds to a candidate, $\varphi$ is the feature map, $\mathbf{w}$ is a *D*-dimensional weight vector with the same size of available data for building up the prediction model, $\varepsilon$ is the noise subject to normal distribution with mean 0 and variance $\zeta$. The random feature map is chosen so that the inner product corresponds to the Gaussian kernel [63]. The open-source Bayesian optimization library COMBO [64] was developed to perform the optimization process automatically.

After the prediction model is constructed, a predictive distribution of transport property is estimated for each remaining candidate. The best candidate is chosen based on the criterion of expected improvement. Finally, the exact transport property is calculated for the chosen candidate, and it is added to the training examples. By repeating this procedure, the calculation of transport property is scheduled optimally, and the best candidate can be found quickly.

As a case study, BO was first applied to design the Si/Ge-composite interfacial structures that minimize or maximize the thermal conductance at room temperature across Si-Si and Si-Ge interfaces as shown in Fig. 4 (a). The interfacial region is composed of 2 unit cells (UC) with 8 Si and 8 Ge atoms with the cross section size of 1 UC × 1 UC, which gives 12,870 possible candidates in total. A binary flag was used to describe the state of each atom: '1' and '0' represent Ge and Si atom, respectively. As for the evaluator, the thermal conductance calculated by atomistic Green's function [46,65,66] is chosen to quantitatively evaluate the performance of each configuration using Atomistix ToolKit simulation package (ATK) [67] with Tersoff potentials [68]. In all calculations, periodic boundary condition was used in the transverse direction (perpendicular to heat conduction direction), and the number of transverse $k$ mesh was set as 20×20, which has been tested to ensure convergence of the transmission calculation.

To test the performance of BO, 10 rounds of optimization were conducted with different initial choices of 20 candidates. As shown in Fig. 4 (b) and (c), all optimizations come to convergence within calculations of 438 structures, which is only 3.4% of the total number of candidates. Insets of Fig. 4 (b) and (c) show the optimal structures obtained by BO for minimum and maximum thermal conductance of Si-Si and Si-Ge interfaces. The optimal Si-Si interfacial structure with maximum conductance is intuitive as the structure provides continuum path of Si for phonons to propagate. However, the other three optimal structures shown are not intuitive and offer new insight. The structures with minimum conductance for both Si-Si and Si-Ge interface were found to be aperiodic superlattices that realize significant reduction from the best conventional periodic superlattice. The optimal structure with maximum conductance for Si-Ge interfaces can be considered as a kind of rough interface, this agrees with the previous AGF calculation result on rough interface [46,69], which showed that the roughness can enhance the phonon transmission at interfaces.

Based on the knowledge learnt above that layered structures give rise to minimum conductance and do not depend on the size of transverse supercell, we performed further optimization of Si/Ge superlattices as shown in Fig. 5: the thickness of the unit layer (UL) is 5.43 Å, and total thickness of interfacial structure ranges from 8 to 16 ULs (from 4.35 nm to 8.69 nm). Similar to the descriptors used in the alloy structure optimization, 8 binary flags were used to indicate the state of each UL ('1' indicates Ge and '0' indicates Si). By performing BO, all the optimal structures can be obtained for Si-Si and Si-Ge interfacial superlattices with different thickness, equal or variable fraction of Si/Ge atoms. It was found that as the layer thickness and number of thickness increase, the thermal conductance decreases and eventually asymptotically converges to a constant value, which is consistent with the trends seen in former investigation of Si/Ge structures [65,70]. When considering a superlattice with a given total thickness, the layer thickness and number of interfaces are two competitive parameters, and this gives rise to the optimal structure with minimum thermal conductance. Another merit of MI lies in possibility to explore new physics in the course of understanding its output. By performing further systematic analyses, it was identified that the small thermal conductance in the aperiodic superlattices originates from their degrees of freedom to mutual-adoptively balance the two competing effects: Fabry–Pérot wave interference [71,72] and interfacial particle scattering [73-75], which reduces the conductance as thickness of the constituent layers in superlattice increases and decreases, respectively. Consequently, the optimal aperiodic structure was found to restrain the constructive phonon interference, making the phonon transport to approach its incoherent limit.

The Bayesian optimization can also be applied to cases which involve multi transport properties, for example the thermoelectric application. The energy conversion efficiency of thermoelectric devices is characterized by the figure-of-merit: $ZT = S^2\sigma T/\kappa$, where $S$, $\sigma$, $\kappa$ are the Seebeck coefficient, electrical conductivity, and thermal conductivity (consisting of electron and phonon contribution) of the material, respectively, and $T$ is the absolute temperature. It has been a challenge to increase $ZT$ through independent control of either electron or phonon properties or, even better, simultaneous improvement of the both, due to the general correlation between the electronic and phononic transport. It has been recently shown that Bayesian optimization is capable of realizing such multifunctional optimization to find nanostructure with maximum $ZT$ among a number of candidates. A successful case study has been demonstrated for defective graphene nanoribbons (GNRs) [76], in which two typical structures are considered, periodically nanostructured GNR and antidot GNR, as shown in Fig. 6.

The multifunctional structural optimization for periodically nanostructured GNR is performed with the Bayesian optimization and its efficiency is compared with random search for different number of removing atoms $m$. The Bayesian search accelerates the exploration of high $ZT$ structures as shown in Fig. 6 (a). In most cases, top-0.5% of the structures can be found by the Bayesian search with half the calculations for the random search. The absence of the $m$-dependence indicates that the efficiency is independent of the total number of candidate structures, which means a comparable efficiency is expected for case of larger GNR systems. Figure 6 (b) compares thermoelectric properties of the pristine structure, the periodic, and the optimal antidot structure. The optimal structure has an aperiodic array of antidots, which increases $ZT$ by 11 times. It is interesting to note that simply arranging the antidot periodically increases $ZT$ by 5.0 times compared with the pristine structure, yet the remaining 2.1 times does require the optimization. This indicates that the optimization of the arrangement of antidots can effectively improve thermal and electronic properties, simultaneously.

When the total number of candidates is relatively larger, BO becomes more memory and time consuming. There are two options in such situation. One is to divide the total number of candidates into small groups and search optimal structure in the sub-groups, and then obtain the final global optimal structure among the local optimal structures from sub-groups. The other is to first gain knowledge in the class of optimal structure based on smaller system. Take the optimization of Si-Ge interfacial alloy structure as example, when the cross section area increases to 2UC × 2UC, searching blindly for all the candidates would explode the number of candidates ($\sim 10^{18}$). We can reduce the number candidates to be in affordable range ($\sim 10^5$) by choosing important subset of candidates following the knowledge learned from smaller system, maintaining the generality as much as possible [61].

**3.2 Monte Carlo Tree Search**

The Bayesian optimization is very effective and accurate when the total number of candidate is around several hundred thousand. However, when dealing with cases with huge or even unlimited number of candidates, it becomes very difficult. Here, we introduce another effective method named Monte Carlo Tree Search (MCTS) [77], which combines the generality of random simulation with precision tree search. MCTS is a popular method for making optimal decisions in artificial intelligence (AI) problems, such as Go games.

The MCTS algorithm is based on a search tree built node by node according to the evaluation of each simulated case, as shown in Fig. 7. Each node contains two important information: an estimated value based on simulation results and the number of times it has been visited. The process of MCTS is composed four typical steps: selection, expansion, simulation, and backpropagation. (i) Selection: Starting at root node R, recursively select optimal child nodes according to larger or small upper confidence bound (UCB) score until a leaf node L is reached. The upper confidence bounds score is calculated by:

$$u_i = \frac{V_i}{n_i} \pm b\sqrt{\frac{2\ln N_{parent}}{n_i}}, \tag{4}$$

where $b$ is a tunable bias parameter to balance the tree exploration and exploitation, $V_i$ is the accumulated simulation values of all structures that was played out from this node through all visits, $n_i$ is the number of the times the node has been visited, $N_{parent}$ is the total number of times that its parent has been visited. + and - are for maximum and minimum conductance optimization case, respectively. (ii) Expansion: If the leaf node L is a not a terminal node then create one or more child nodes and select one node C. (iii) Simulation: Randomly select one playout from node C and do the conductance calculation. (iv) Backpropagation: Use the calculated thermal conductance value to update the $n_i$, $N_{parent}$ and $V_i$ values of the nodes on the path back from node C to node R. It has to be mentioned that MCTS does not guarantee finding global optimal structure, and instead it offers structure close to the global optimal one with high efficiency.

The MCTS was applied to Si/Ge interfacial alloy system to test the performance. The convergence of MCTS shown in Fig. 8 is slower compared with BO. Not all the 10 rounds of optimization can target the global optimal structures with the same number of calculated candidate structures as BO, however, they are approaching the global optimal conductance. The advantage of MCTS as summarized in Table 1 is that it can deal with optimization cases with huge or unlimited number of candidates that BO cannot deal with. With the increase of number of candidates, the consumed time for selection of next candidate in BO will increase quickly, which make the BO optimization rather time consuming, while the MCTS is able to obtain the quasi-optimal structures with high efficiency. As a case study, we applied MCTS to optimize the interfacial Si-Ge roughness [78] as shown in Fig. 9 (a), in which we divide the design region to 10 tree layers, the thickness of each layer is 5.43 Å, and each node in the tree has four child (0, 1, 2, 3), this gives the total number of candidates of 1,048,576. As shown in Fig. 9 (b), within around 300 structures calculation, the interfacial thermal conductance quickly increases from 377 MWm$^{-2}$K$^{-1}$ to 408 MWm$^{-2}$K$^{-1}$. Figure 9 (c) shows the interfacial thermal conductance versus the roughness which is defined by the real rough surface area divided by the projected area along the phonon transport direction. The result indicates that the maximum interfacial thermal conductance appears in the middle value of interfacial roughness, which is not intuitive. By comparing the phonon transmission of the optimal and flat interface shown in Fig. 9 (d), we can find that the transmission in the middle frequency range from 4 to 10 THz is obviously enhanced.

## 4. Summary and perspectives

There have been successful applications of MI on heat transfer problems during the past decade. MI has been able to discover crystals with ultimate high/low thermal conductivity, alloy structures with maximum/minimum thermal conductance, superlattices with minimum thermal conductance, defective graphene nanoribbons with highest thermoelectric figure of merit, and rough interfaces with high thermal conductance. MI not only discovers optimal structures efficiently, but also help us understand new physics behind the found/designed novel materials and structures. It is also worth mentioning that the MI-based design algorithm can be easily extended to transport of other quasi particles (e.g. electron, photon, magnon).

Of course, there are still several technical challenges to be overcome as listed below.

(1) There is large gap between the "big data" required for credible machine learning and the "small data" we can collect. Take the database screening for crystals with high/low thermal conductivity as an example. We can collect "big data" for harmonic phonon property in short period, but only "small data" for the thermal conductivity due to the heavy calculation of anharmonic property. Here one possible solution is the transfer learning, which obtains a prediction model from a small database by partially or initially adopting the model parameters trained for related properties with larger database.

(2) Current structure design is still limited to small scales and systems, considering that many materials of interests in the context of thermal transport consists of multiscale structures ranging from nanometer to micrometers. When scaling up the computed system, one challenge during is the calculation time of thermal transport properties. Note that in enlargement of the system may also mean involvement of more physics, which can make the calculation itself more complex and heavy. For instance in the case of above-discussed layered structure, as the layer thickness increase, phonon-phonon scattering becomes important, which requires calculations with anharmonic lattice dynamics that is more expensive than the harmonic AGF calculations. Since optimization typically involves calculations of thousands of structures, if calculation of each candidate structure takes several hours, the entire calculation become very expensive. Therefore, improvement the calculation speed and efficiency is directly related to usability of the optimization scheme. Another option is to balance the accuracy and speed to an acceptable level to save the calculation time and improve the designing efficiency. This can also be done in hierarchical manner, where the first screening is done with low accuracy but followed by refining step with higher accuracy.

(3) The development of new effective descriptors is important and necessary. For the database screening, the combined structure and chemical descriptors are mainly used for screening crystals with ultimate high and low thermal conductivity. The descriptor does not necessarily need to be a physical property, and abstract quantities such as radial distribution function has been shown to perform as well or often even better. Physical properties on the other hand is powerful when taking advantage of the known physical correlation. For instance, in an on-going work, we have adopted the phonon scattering phase space (harmonic property) as the feature descriptor to search for high thermal conductivity crystals, which can avoid the heavy calculation of anharmonic properties and thus greatly reduce the computational load [45]. For the nanostructure designing, the current descriptors are binary values denoting the element kind, which has merit of being intuitive to use, but there should be better descriptors to represent the structure, for instance incorporating also the combination of neighboring elements. Development of new descriptors are underway to further improve the efficiency of the optimization.

(4) It is difficult to know the efficiency of the search or optimization a priori. The search efficiency dependents on each specific case, especially on the feature of the histogram of the transport properties for all the candidates. The Bayesian optimization was around 3.4% for the Si/Ge alloy structure case presented above but this may change for other cases. One could perform an optimization test first with a smaller system, and judge the suitability of a specific method to the system of such kind, although this certainly assumes that similar class of structures result in similar efficiency, which of course cannot be generalized.

(5) There is often some gap between the optimal design obtained and the actual realization of the material. Take the non-periodic superlattice structure designing as an example, the design so far has been purely computational, and is expected to be followed by fabrication. However, the actual fabrication involves practical issues such as the non-sharp interface due to inter-diffusion at the interface, and the error in each layer thickness due to uncertainty in the growth/deposition speed. These structural deviations, when they are significant, can certainly spoil the optimization. One possible approach is to incorporate the uncertainties in fabrication process into the evaluation function in the optimization. Another is to directly combine MI with experiment i.e. to perform optimization using also or only experimentally measured data, output of which is the actual material instead of a design.

Besides the above listed challenges, there is also some non-technological issues to solve, for example, the first-principles based ALD calculation has become a standard tool for thermal conductivity calculation, but the results from different researchers or groups are not well organized. A challenge that can certainly be overcome is to bridge the difference among input and output formats of various tools used for the calculation such as VASP [79,80], Quantum ESPRESSO [81], etc. for the DFT force calculations, and ALAMODE [82], Phonon3py [83], ShengBTE [84], etc. for the ALD calculations. Besides, there are issues such as the setting parameters during the calculation as the mesh size, cut off length, pseudopotentials etc., and thus, standardization of calculation settings and validation procedure would be necessary.

There is an opportunity to build up a standard database for collecting and sharing ALD data from different individuals.

In any case, in the near future the application of MI to heat transfer is expected to expand in various forms of heat transfer (conduction, convection and radiation), from nano to macro scales, and from simulations to experiments.

**Acknowledgments**

We thank the collaborations with Koji Tsuda, Thaer M. Dieb, Kasuki Yoshizoe, Ryo Yoshida, Kenta Hongo, Terumasa Tadano, Takuma Shiga, Lei Feng, Zhufeng Hou, Masaki Yamawaki and Masato Ohnishi. This work was supported by "Materials Research by Information Integration" Initiative (MI$^2$I) project of the Support Program for Starting Up Innovation Hub and CREST Grant No. JPMJCR16Q5 from Japan Science and Technology Agency (JST), and KAKENHI Grants No. 16H04274 from Japan Society for the Promotion of Science (JSPS).

**Figures and captions**

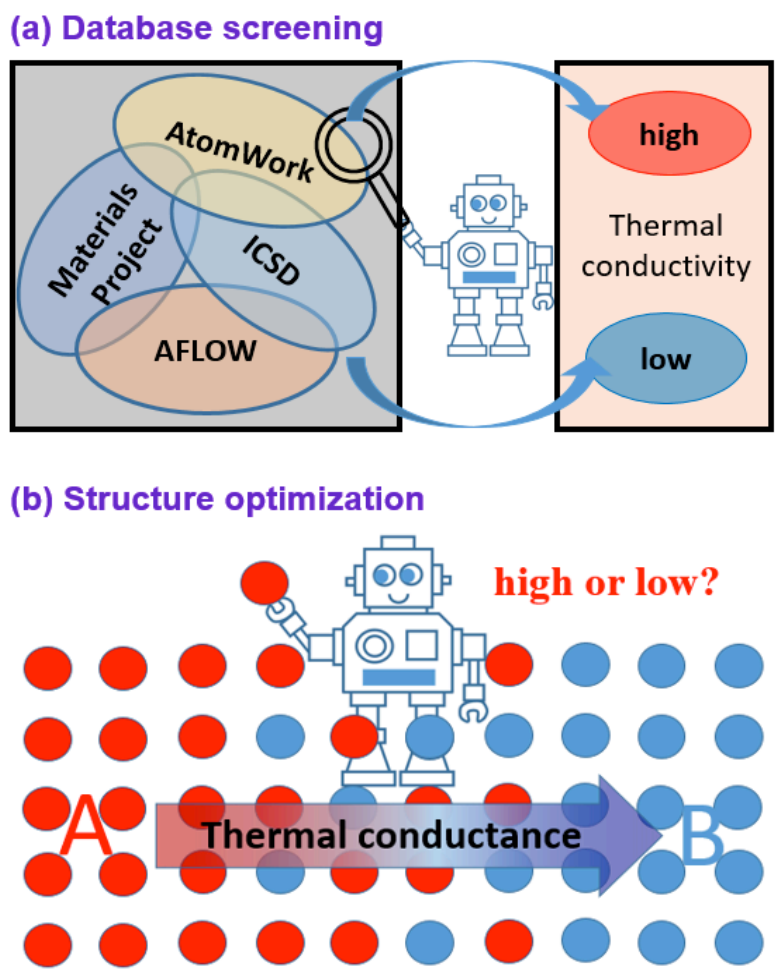

**Fig. 1.** Application of materials informatics on thermal transport: (a) database screening, (b) structure optimization.

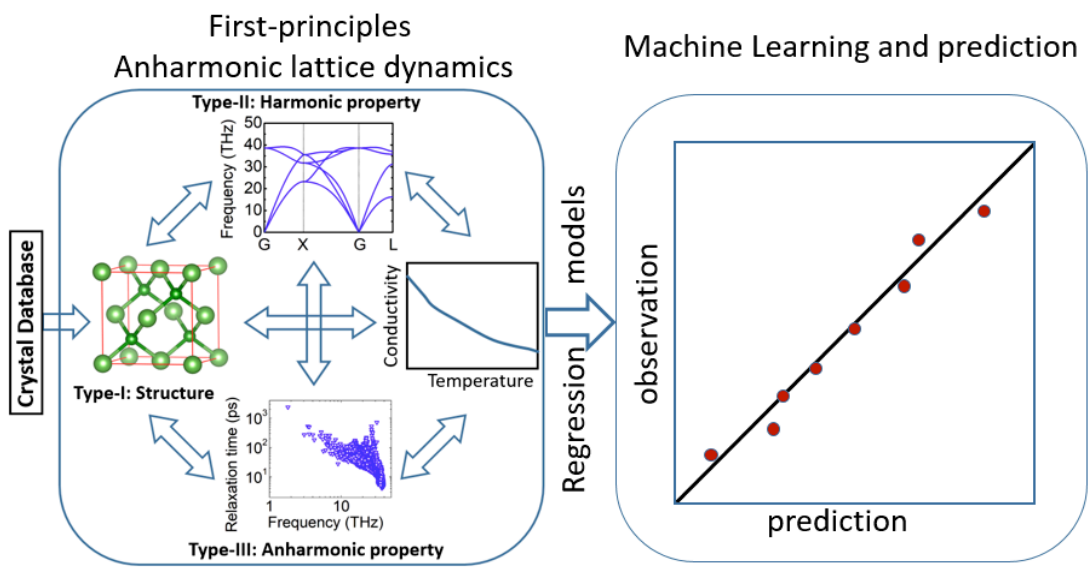

**Fig. 2.** Schematics of the high throughput screening by combining the first-principles anharmonic lattice dynamics and machine learning. The descriptors currently widely used for machine learning are structure and chemical information.

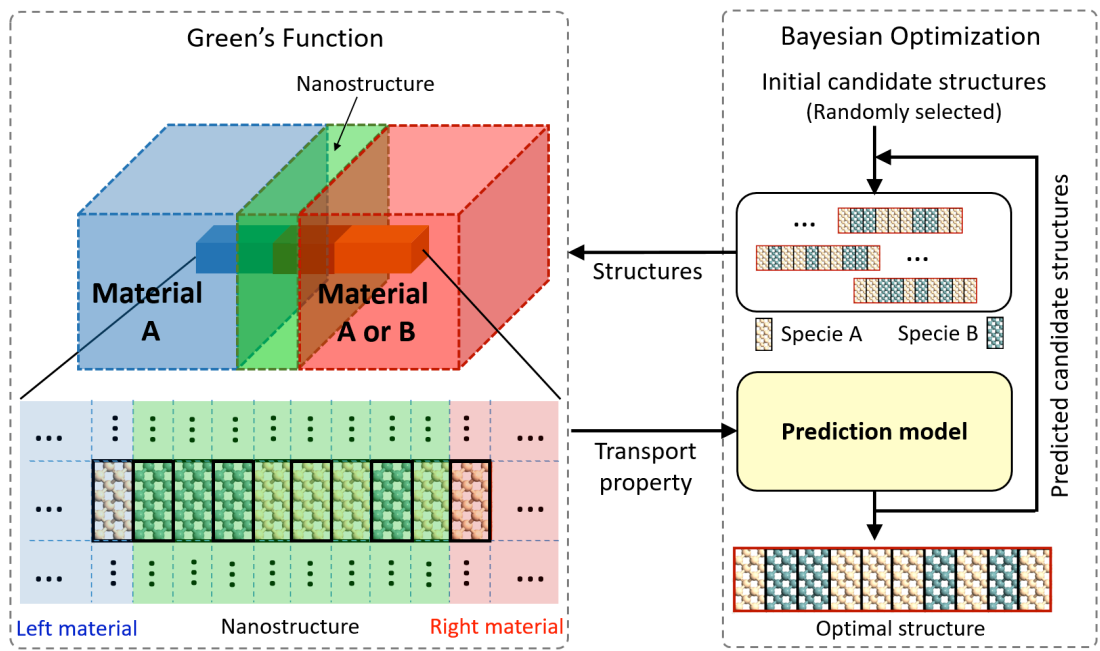

**Fig. 3.** Schematics of the Bayesian optimization combing with the atomistic Green's function for transport property calculation.

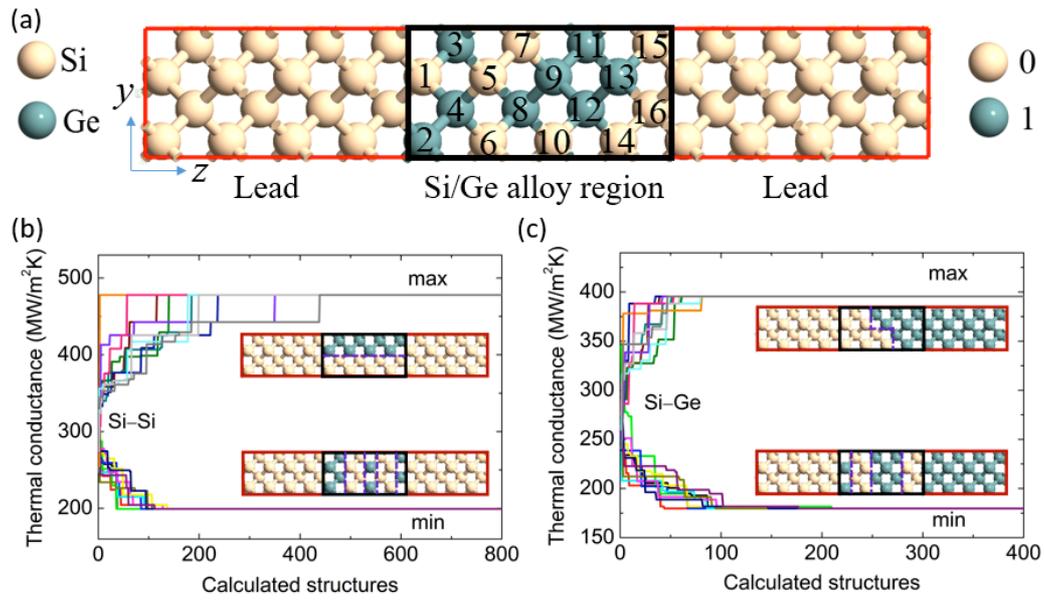

**Fig. 4.** Bayesian optimization of interfacial Si/Ge alloy structure for maximum and minimum thermal conductance. (a) system for atomistic Green's function calculation, (b) and (c) show the 10 optimization runs with different initial choices of candidates for Si-Si and Si-Ge cases, respectively. The insets show the coeesponding optimal structures for maximum and minimum thermal conductance.

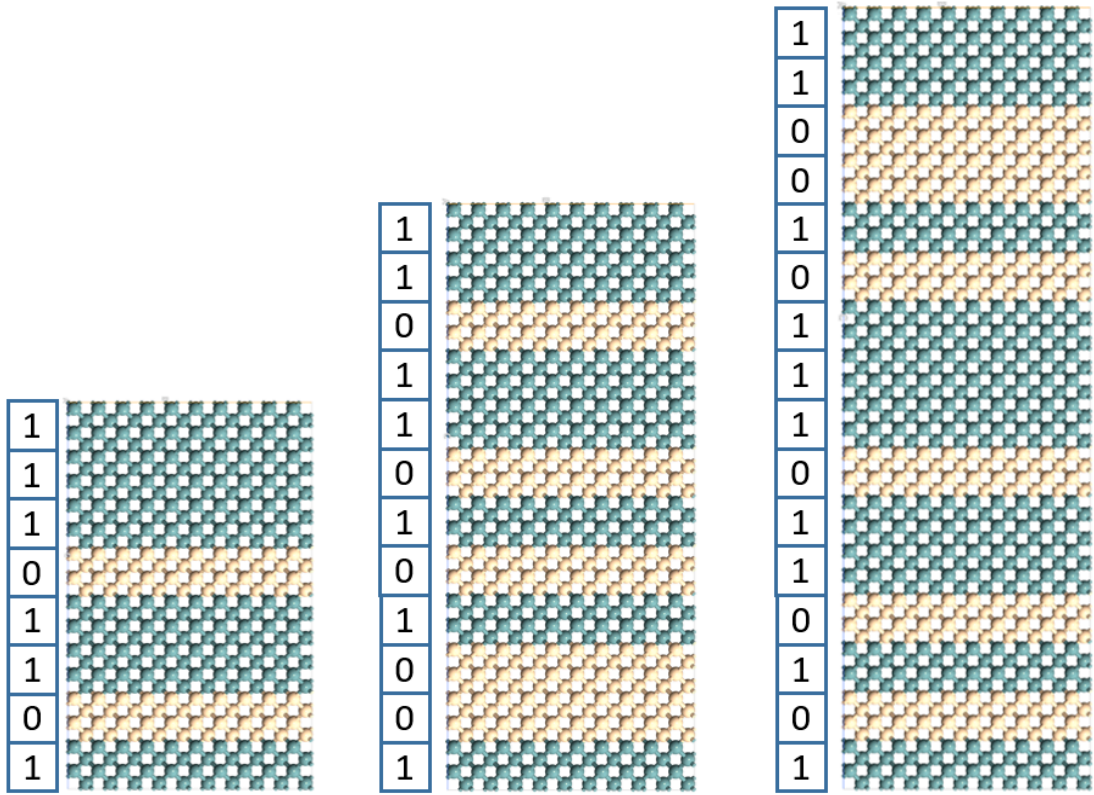

**Fig. 5** Designed non-periodic Si/Ge superlattice with minimum thermal conductance for a given total thickness. The obtained thermal conductance is siginificantly lower than the correspnoding periodic superlattice.

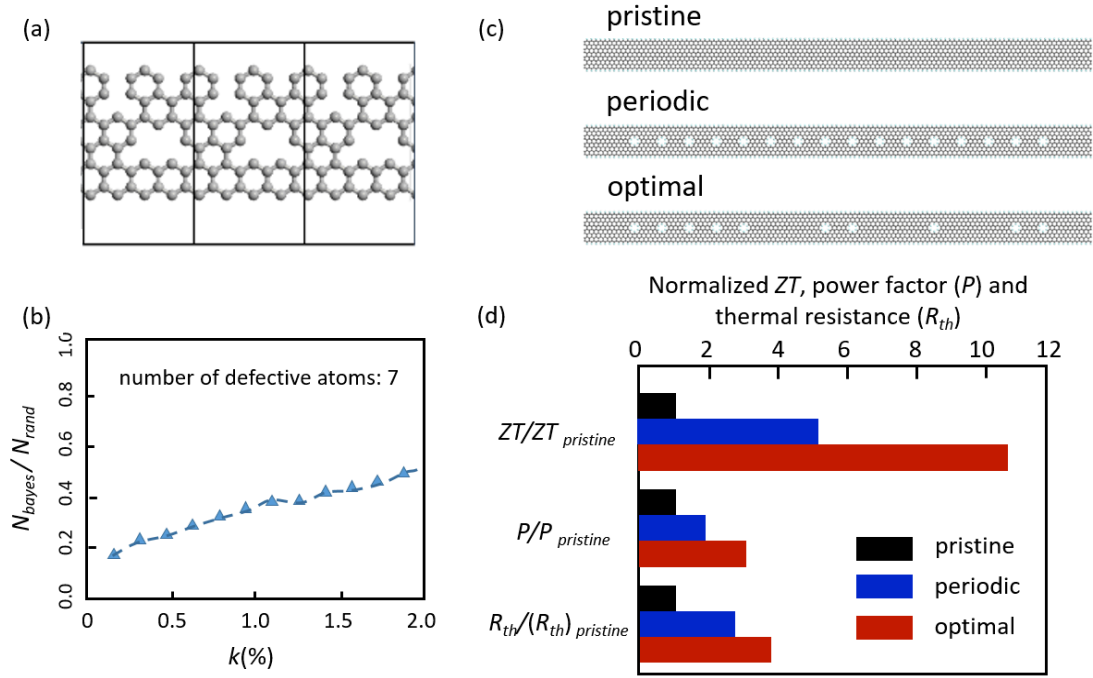

**Fig. 6** Designing defective graphene nanoribbons for thermoelectric application using Bayesian optimization. (a) Optimization of periodically nanostructured graphene nanoribbon. (b) Comparision of Bayesian optimization and random search by the average number of calculations needed until a structure that belongs to the top $k$% of all the candidates. (c) Optimziation of antidot graphene nanoribbon. (d) Comparision of the normalized $ZT$, power factor ($P$) and thermal resistance ($R_{th}$) for the pristine, periodic and coptimal structures.

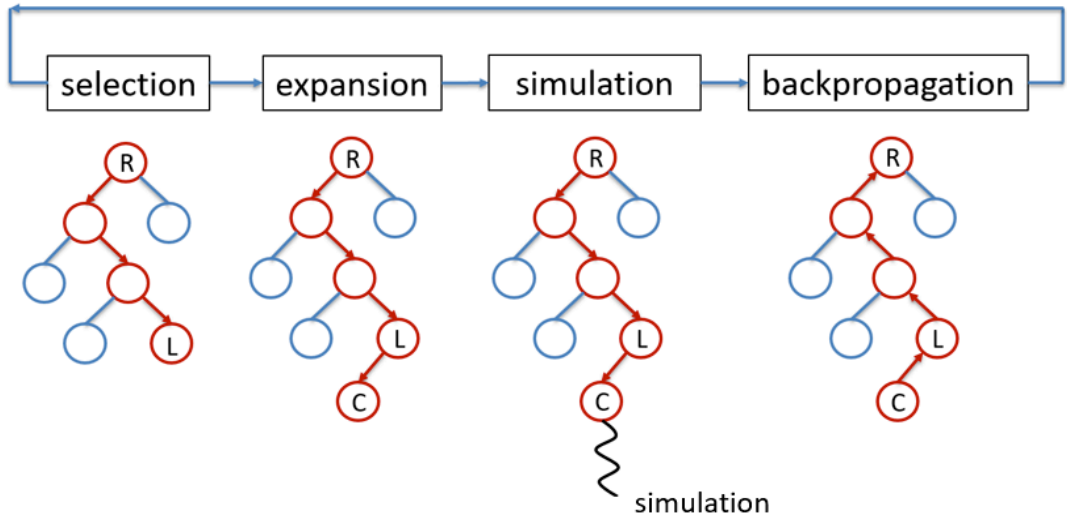

**Fig. 7** Schematics of the Monte Carlo tree search (MCTS) algorithm, which is composed of four processes: selection, expansion, simulation, and backpropagation. The R, L and C are nodes in the tree.

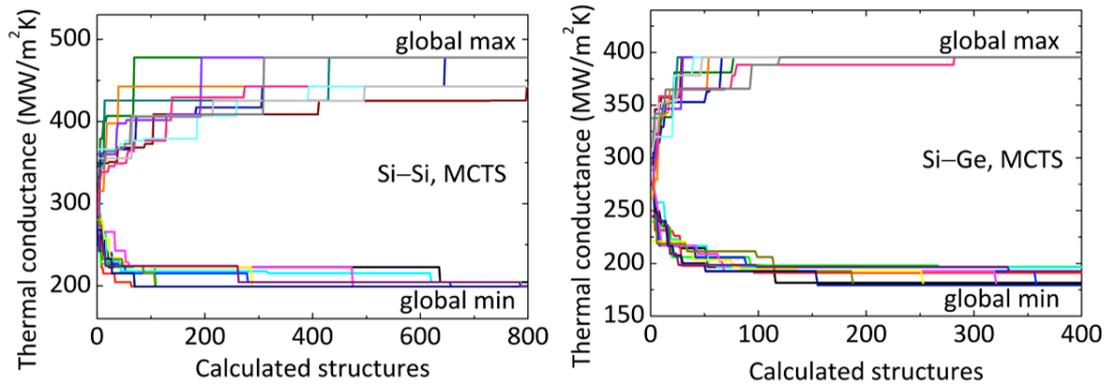

**Fig. 8** Performance of Monte Carlo tree search for Si-Si and Si-Ge interfacial alloy structure optimization.

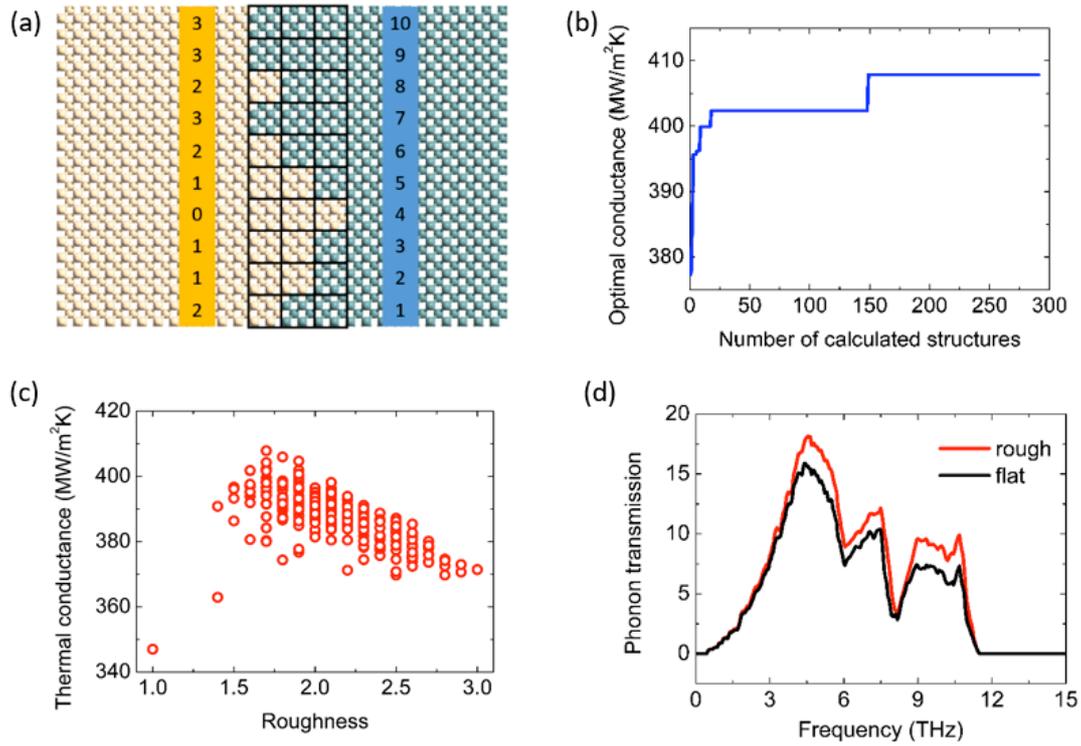

**Fig. 9** (a) Designing rough interfacial structure by Monte Carlo tree search, (b) Performance of Monte Carlo tree search optimization, (c) Thermal conductance versus the roguness, (d) Comparision of the phonon transmission of optimal rough and flat interfacial strucutre.

**Table 1.** Comparison of Bayesian optimization and Monte Carlo tree search.

| Informatics method | | Bayesian optimization (BO) | Monte Carlo tree search (MCTS) |
|---|---|---|---|
| Candidates number | | limited (<200,000) | huge or unlimited |
| Candidates preparation | | all listed | Automaticlaly generated |
| Optimization time | | ∝ candidate number | very short |
| Efficiency | <200,000 | high | lower than BO |
| | >1,000,000 | Low or failed | higher than BO |